\documentclass[preprint,samsmath,amssymb,floatfix,superscriptaddress]{revtex4}
\usepackage{graphicx}
\usepackage{dcolumn}
\begin{document}
\title{
An operational scheme to determine the locally preferred structure
of model liquids}
\def\paris{\affiliation
{Laboratoire de Physique Th\'eorique de la Mati\`ere Condens\'ee,
Universit\'e Pierre et Marie Curie, 4 place Jussieu, 
Paris 75005, France}}
\def\esrf{\affiliation{
European Synchrotron Radiation Facility, B.P. {\em 220},
F-{\em 38043} Grenoble, Cedex France}}
\author{S.~Mossa}\esrf
\author{G.~Tarjus}\paris 
\date{\today}
\begin{abstract}
We present an operational method to determine the 'locally preferred structure'' 
of model liquids, a notion often put forward to explain supercooling of a liquid
and glass formation. The method relies on finding the global minimum in the (free) 
energy landscape of clusters of atoms or molecules embedded in a liquid-like 
environment. We propose a more systematic approach of the external potential 
mimicking the influence of the surrounding bulk liquid on the cluster than in our 
previous work [S. Mossa and G. Tarjus, J. Chem. Phys {\bf 119}, 8069 (2003)]. 
The procedure is tested on the one-component Lennard-Jones liquid and we recover, 
without a priori input, that the locally preferred structure is an icosahedral 
arrangement of thirteen atoms.  
\end{abstract}
\maketitle
An intuitive idea that has been put forward for explaining both supercooling of liquids
below their melting point~\cite{frank52} and glass formation~\cite{nelson89,kivelson95}
is that a liquid is characterized by a {\em locally preferred structure}
which possesses two important properties: it is different than the local 
arrangement of the atoms or molecules in the crystal and it cannot propagate
freely in space to generate a periodic structure with long-range order.
The former accounts for the local structural reorganization (and the associated energy cost)
that occurs in a first-order crystallization; the latter, which is known as 
geometric or structural frustration~\cite{sadoc99}, explains why, despite the local 
preference for a different arrangement, the crystalline structure is globally 
preferred below some temperature.

Although intuitive and plausible, the concept of locally preferred structure
in liquids needs a firm footing. The most studied, and often quoted, example
of such a locally preferred structure is the local icosahedral order in atomic liquids
in which the atoms interact through spherically symmetric potentials (as the one-component
Lennard-Jones model). Frank indeed showed, more than fifty years ago~\cite{frank52},
that the lowest-energy cluster made by 12 atoms around a central one is an
icosahedron rather than crystalline local arrangements associated with 
face-centered cubic and hexagonal close-packed lattices.
This propensity to form icosahedral or, more generally, polytetrahedral structures has been 
since confirmed, both experimentally~\cite{filipponi99,reichert00,shenk02,dicicco03} 
and numerically~\cite{steinhardt83,nelson89,dzugutov02,doye03}. However, even this simple 
example is not devoid of problems: for small isolated clusters, the energetics, 
and correspondingly the relative stability of the different configurations, 
is dominated by surface effects coming with the fact that the outermost atoms of an 
isolated cluster have a reduced number of interacting neighbors compared to 
the inner atoms or to atoms in bulk liquid.
(Even in an isolated cluster of 1000 atoms, more than 30$\%$ of the atoms are 
at the surface.) One may then wonder if the preference for icosahedral order in a 13-atom cluster
does not merely result from such surface effects. Note that one cannot simply cure this problem
by resorting to the application of periodic boundary conditions. In this situation
indeed, only the local arrangements that can be easily extended to the whole space 
by periodic replication can reasonably be found as ground states: frustrated local structures
cannot be selected in this way.

On general grounds, a locally preferred structure in a liquid should be a 
configuration of atoms or molecules that minimizes some ``local free energy''. 
In most cases, one expects such a structure to be very difficult to determine 
in experiments because its characterization requires information about many-atom
correlations, beyond the usual pair correlations~\cite{steinhardt83}.
Numerical studies on model liquids are thus for now the only practical means, 
but an operational definition of the appropriate ``local free energy'' is required.
We have recently proposed a method which consists of searching for the ground state 
of clusters of a given number of atoms placed in a cavity and subject to an
external field that mimics the interaction with the rest of the liquid~\cite{mossa03}.
The structure of the outside liquid at the chosen density and temperature
only enters the calculation through the bulk pair distribution function
(known from simulation studies). The influence of the liquid on the cluster
is thus treated at a mean-field level, with the two-fold advantage
of minimizing the surface effects by embedding the cluster in a 
condensed liquid-like phase and minimizing the frustration effects by neglecting
the detailed many-body structure of the environment.
One can thus avoid the ``two evils of finite cluster geometry or periodic replication''
discussed by Hoare~\cite{hoare78}.   

Applied to a single-component Lennard-Jones system, the method has allowed us 
to confirm that the icosahedral arrangement of $13$ atoms is the locally 
preferred structure of the liquid~\cite{mossa03}. However, this first implementation 
has a shortcoming. The cavity in which an $N$-atom cluster is confined is defined 
by a sharp cut-off, $R_C=r_{max}+\mu\sigma$, where $r_{max}$  is the distance of the 
outermost atom of the cluster to the center of the cavity, $\sigma$ is the 
Lennard-Jones atomic diameter, and $\mu$ is a constant chosen to account for the fact
that repulsive interactions between atoms make very unlikely the presence of
``bulk-liquid atoms'' whose centers are too close to those of the cavity atoms~\cite{mossa03}
(typically , $\mu\simeq 0.5$).
The procedure for finding the global energy minimum of an $N$-atom cluster in the 
external field generated by a liquid environment uses an optimization 
algorithm~\cite{wales97,doye98} adapted to minimize the total energy also with respect 
to the cavity radius $R_C$. The sharp cut-off defined above leads to the annoying
feature that the minima of the ``(free) energy landscape''~\cite{note}, i.e.,
of the hypersurface formed by the sum of the intra-cluster Lennard-Jones interaction energy
and of the external potential due to the outside liquid environment plotted as
a function of the configurational coordinates of the atoms of the cluster, may 
correspond to ``cusps'' in a few directions on top of the usual analytic characterization 
(first derivative equal to zero and positive curvatures) in the other directions.
The Hessian matrix, formed by the second derivatives of the energy function, is thus
ill-defined, which for both technical and physical reasons is an unwanted limitation.

In this note, we propose a way out of the above problem and check the relevance of 
our procedure to the same one-component Lennard-Jones model studied before.
We also discuss the more interesting generalization to molecular liquids.

We consider a system of atoms interacting via a pairwise additive spherically symmetric 
Lennard-Jones potential $v(r)=4 \epsilon \left(\left( \sigma/r\right)^{12}-
\left( \sigma/r\right)^{6} \right)$. A number $N$ of atoms are placed in a cavity $\cal{C}$
of radius $R_C$ that we envisage as surrounded by bulk liquid 
made of the same atoms and characterized at the chosen temperature $T$ and density $\rho$
by a known pair distribution function $g(r;T,\rho)$.
This latter is assumed to be unaffected by the existence of a cavity
and the outside liquid is essentially considered as a continuum as in a mean-field
type of description. However, one knows from standard liquid-state theories~\cite{hansenbook}
that some care is needed in applying mean-field approaches; the short range 
repulsive interactions cannot be treated at a mean-field level and should rather be
included in a reference system described as exactly as possible.
Accordingly, when considering the interaction between cavity atoms and outside liquid
one should treat differently the attractive component, which can be simply 
described by a mean-field type of approximation, and the short-range repulsive
one, which accounts for the exclusion effect creating the cavity.

\begin{figure}[t]
\centering
\includegraphics[width=0.65\textwidth]{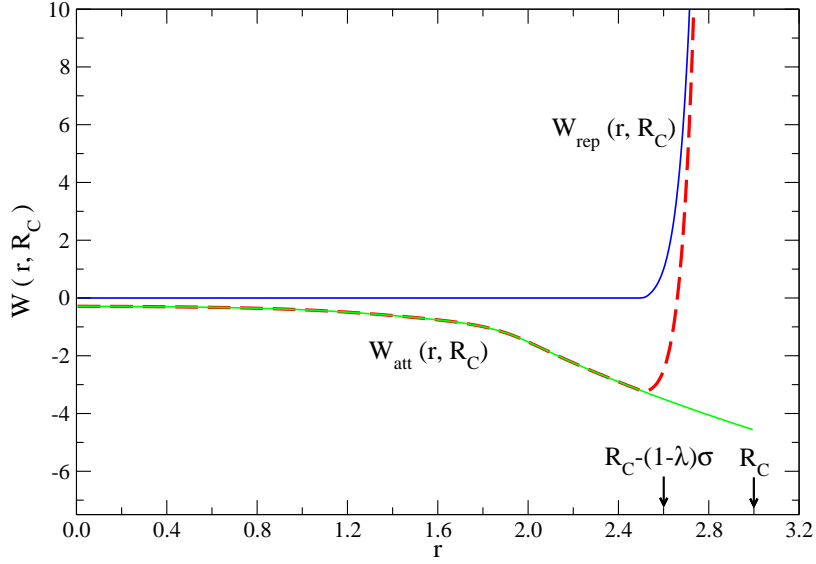}
\caption{External potential due to a liquid environment acting on an atom 
placed  in a cavity of radius $R_C$ at a distance $r$ from the center.
The attractive and repulsive contributions are shown as full lines and the sum 
as a dashed line. ($R_C=3 \sigma$, $\lambda =0.6$, $\rho =0.88$, $T=1.095$).
The repulsive component diverges for $r=R_C+\lambda\sigma$ and the
minimum of the full potential is close to $R_C-(1-\lambda)\sigma$.}
\label{fig:pot}
\end{figure}

A convenient starting point is given by the Weeks-Chandler-Andersen~\cite{weeks71}
splitting of the Lennard-Jones interaction into a purely attractive piece,
$v^{WCA}_{att}(r)=v(r)$ for $r\ge r_{min}$ and $v^{WCA}_{att}(r)=-\epsilon$ 
for $r\le r_{min}$, $r_{min}=2^{1/6}\sigma$ being the location of the minimum of the 
Lennard-Jones potential, and a purely repulsive piece  $v^{WCA}_{rep}(r)=v(r)-v^{WCA}_{att}(r)$
that is zero for $r\ge r_{min}$. The mean-field-like attractive potential energy
acting on a given atom of the cluster at position ${\bf r}$ due to the liquid outside a
cavity  of radius $R_C$ is then given by
\begin{equation}
W_{att}(r ;R_C)=\frac{\rho}{2}\int_{|{\bf r' }|>R_C} 
d^3 {\bf r'}\; g(|{\bf r}-{\bf r'}|; T,\rho) \; v^{WCA}_{att} (|{\bf r}-{\bf r'}|),
\label{meanfield:eq}
\end{equation}
which can be rewritten as a simple one-dimensional integral~\cite{mossa03}
and calculated, as well as its first and second derivatives, from the known 
$g(r;T,\rho)$~\cite{matteoli95}. This attractive mean-field potential must
be supplemented by a repulsive potential that confines the cluster inside the cavity
and guarantees the absence of strong overlap between cavity atoms and ``bulk-liquid
atoms''. An effective description of this confinement effect is provided by simply
considering a spherically symmetric repulsive potential around the center of the cavity:
\begin{equation}
W_{rep}(r;R_C)=v^{WCA}_{rep}(R_C+\lambda\sigma-r),
\label{conf:eq}
\end{equation}
where $\lambda$ is a constant chosen to avoid too severe overlaps between
atoms in the cavity (other than the one at position ${\bf r}$) and ``bulk-liquid
atoms''. The sharp cut-off used in our previous work~\cite{mossa03}amounts to
replace $v^{WCA}_{rep}$ by a hard-sphere interaction and choosing $\mu=1-\lambda$.
The shape of the total external potential and of its two,
attractive and repulsive, contributions is illustrated if Fig.~\ref{fig:pot}.
(For purely technical reasons and with no detectable consequences,
we have smoothed the WCA repulsive interaction around $r_{min}$ to ensure
a finite, albeit very large second derivative.)

\begin{figure}[b]
\vspace{1.0cm}
\centering
\includegraphics[width=0.65\textwidth]{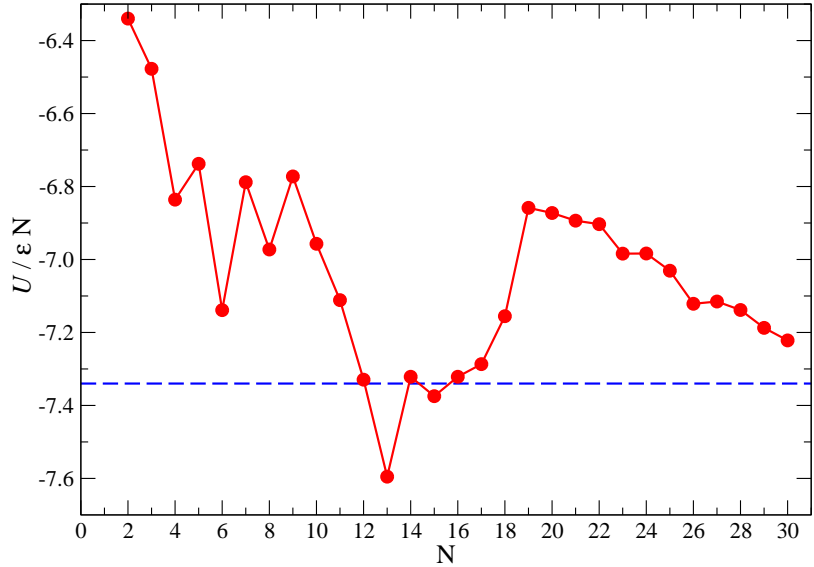}
\caption{Energy per atom of the ground state of $N$-atom clusters
embedded in a liquid-like environment. The horizontal line
is the corresponding bulk energy per atom. Note the clear minimum
at $N=13$.}
\label{fig:ene}
\end{figure}

With the external potential so defined, one can determine the ground-state configuration
of an $N$-atom cluster in a liquid-like
environment (at given $T$ and $\rho$) by searching the global minimum of the total energy
\begin{equation}
{\cal U}\left(\left\{{\bf r}_j\right\}_{1,\ldots,N};R_C\right)=
\sum_{i<j=1}^N v(|{\bf r}_i-{\bf r}_j|)
+\sum_{j=1}^N 
(W_{att}(r_j;R_C)+W_{rep}(r_j;R_C))
\label{U_tot:eq}
\end{equation}
with respect to variations of the positions of the $N$ atoms and of
the cavity radius. The minimization procedure is no longer hampered by ``cusp'' 
problems and the Hessian matrix is easily calculated. 
The resulting energy per atom for clusters of sizes between $N=2$ and $N=30$ is 
shown in Fig.~\ref{fig:ene} for a typical state point near the triple point of the system
($\rho=0.88$, $T=1.095$ in usual reduced Lennard-Jones units) and $\lambda=0.6$
(which in all cases leads to $R_C-r_{max}\simeq 0.5\sigma$).
The result comes with no real surprise: the locally preferred structure
is obtained for $N=13$ atoms and the configuration (not shown here) is a perfect icosahedron.
We have also directly checked that the energy per atom of the corresponding crystalline 
cuboctahedral arrangement of $13$ atoms is higher.

The main conclusion is that we have now an operational method for determining the locally 
preferred structure of model liquids. An important property of our method is that a direct comparison 
of the energy per atom for clusters of different sizes is meaningful:
accordingly, no a priori knowledge of the characteristics of the putative locally
preferred structure (symmetry, number of atoms or molecules in the local arrangement)
is required. Considering an atomic, single-component system as done here and in our previous work 
is a necessary check of the procedure, because a priori knowledge about
local atomic arrangements does exist. More interesting however will be the study of
molecular glassforming liquids. Preliminary results on the Lewis and Wahnstr\"om
model of ortho-terphenyl~\cite{lewis94} seems to support the applicability of the 
present method~\cite{mossanext}.   
\newpage

\begin{thebibliography}{99}
\bibitem{note}
Due to the mean-field contribution generated by the liquid environment,
one should rather speak of a ``free-energy landscape'' than of an 
``energy landscape'': the contribution of the outside liquid has been already
averaged and depends on temperature and density via the pair distribution function.
\end{thebibliography}
\end{document}